\documentclass[aps,prl,preprint,superscriptaddress]{revtex4-1}
\usepackage[latin1]{inputenc}	
\usepackage[T1]{fontenc}
\usepackage[cyr]{aeguill}
\usepackage[pdftex]{graphicx}
\usepackage{wrapfig}
\usepackage[english,french]{babel}
\usepackage{amsmath}
\usepackage{amssymb,amsfonts,textcomp}
\usepackage{color}
\usepackage{array}
\usepackage{hhline}
\usepackage{hyperref}
\usepackage{fancyhdr}
\usepackage[pdftex]{graphicx}
\definecolor{violet}{rgb}{0.5,0,0.5}
\definecolor{vert}{rgb}{0,0.65,0}
\begin{document}

%%%%%%%%%%%%%%%%%%%%%%%%%%%%%%%%%%%%%%%%%%%%%%%%%%%%%%%%%%%%%%%%%%%%%%%%%%%%%%%%%%%%%%%%%%%%%%%%%%%%%%%%%%%%%%%%%%%%%%%%%%%%%%%%%%%%%%%%%%%%%%%%%%%%%%
%%%                                                                  TEXTE :                                                                  %%%
%%%%%%%%%%%%%%%%%%%%%%%%%%%%%%%%%%%%%%%%%%%%%%%%%%%%%%%%%%%%%%%%%%%%%%%%%%%%%%%%%%%%%%%%%%%%%%%%%%%%%%%%%%%%%%%%%%%%%%%%%%%%%%%%%%%%%%%%%%%%%%%%%%%%%%

\thispagestyle{empty}
\title{Electrical Control of the Superconducting-to-Insulating Transition in Graphene/Metal Hybrids}
\author{Adrien Allain}
\affiliation{Institut N\'eel, CNRS-UJF-INP, BP 166, 38042 Grenoble cedex 9, France.}
\author{Zheng Han}
\affiliation{Institut N\'eel, CNRS-UJF-INP, BP 166, 38042 Grenoble cedex 9, France.}
\author{Vincent Bouchiat}
\affiliation{Institut N\'eel, CNRS-UJF-INP, BP 166, 38042 Grenoble cedex 9, France.}

\date{\today}
\maketitle

\clearpage

\textbf{Graphene~\cite{Geim2007}  is a sturdy and chemically inert material exhibiting an exposed two-dimensional electron gas of high mobility. These combined properties enable the design of graphene composites, based either on covalent~\cite{Elias2009} or non-covalent~\cite{Kozlov2011} coupling of adsorbates, or on multilayered structures~\cite{Ponomarenko2011}. These systems have shown novel tunable electronic properties such as bandgap-engineering~\cite{Kozlov2011}, reversible metal/insulating transition~\cite{Elias2009,Ponomarenko2011}  or supramolecular spintronics~\cite{Candini2011}. Tunable superconductivity is expected as well~\cite{Uchoa2007}, but experimental realization is lacking. Here, we show experiments based on metal/graphene hybrid composites, enabling the tunable proximity coupling of an array of superconducting nanoparticles of tin onto a macroscopic graphene sheet. This material allows a full electrical control of the superconductivity down to a strongly insulating state at low temperature. The observed gate control of superconductivity results from the combination of a proximity-induced superconductivity generated by the metallic nanoparticles array with the two-dimensional and tunable metallicity of graphene. The resulting hybrid material behaves, as a whole, like a granular superconductor showing universal transition threshold and localization of Cooper pairs in the insulating phase.  This experiment sheds light on the emergence of superconductivity  in inhomogeneous superconductors, and more generally, it demonstrates the potential of graphene as a versatile building-block for the realization of novel superconducting materials.}

Even though intrinsic superconductivity in doped~\cite{Uchoa2007} graphene has been proposed, it has not yet been experimentally shown. However graphene can carry supercurrent by means of proximity effect~\cite{Heersche2007} , in which charge carriers in a non-superconducting material acquire superconducting correlations in the vicinity of a superconductor. In the present study, the proximity effect in graphene is not locally generated from the contacting electrodes, like it is done in most mesoscale experiments~\cite{Heersche2007}, but rather by coupling its surface to a macroscopic 2D network of superconducting clusters~\cite{Feigel'man2008}. As opposed to covalent functionalization of graphene, which strongly affects the density of state, non-covalent coupling of adsorbates on graphene is useful for designing materials with novel electronic functions, since the graphene keeps most of its exceptional electronic properties, while gaining others coming from the coupling elements (Fig.1). Previous experiments involving decoration of metal clusters on exfoliated graphene have shown~\cite{Kessler2010} that such a hybrid system can exhibit a gate-tunable Berezinsky-Kosterlitz-Thouless (BKT) transition towards a fully two-dimensional superconducting state with a critical temperature related to the normal state resistivity. In that experiment, the mean free path and the superconducting coherence length exceeded the average gap width separating neighboring tin nanoparticles, resulting in homogeneous 2D superconductivity. Here, the main and crucial difference is the use of centimeter-scale graphene layers grown by chemical vapor deposition (CVD) instead of previously used micron-scaled exfoliated graphene. 

The CVD grown graphene used in this study have a significant electronic disorder which induces strong electron localization under 1K (see characterization of bare graphene samples in Supplementary Figure S1 and S3 and Supplementary Table S1). This results in a completely different behavior : upon gating the tin decorated sample, one observes a superconductor to insulator transition (SIT). The SIT in 2D films of superconductors has been a very active field of research in the condensed matter community for the last twenty years~\cite{Goldman1998}. It is of interest for the study of quantum phase transitions~\cite{Sondhi1997}, but also to understand how superconductivity emerges in high-T$_c$ superconductors~\cite{Bollinger2011}. Recent advances in this field have involved the use of electric field to tune the SIT at constant disorder~\cite{Parendo2005, Bollinger2011,Leng2011}.

\bigskip

\begin{figure}[htbp]
\begin{center}
\hspace*{-0.5in}
	\includegraphics[width=17cm]{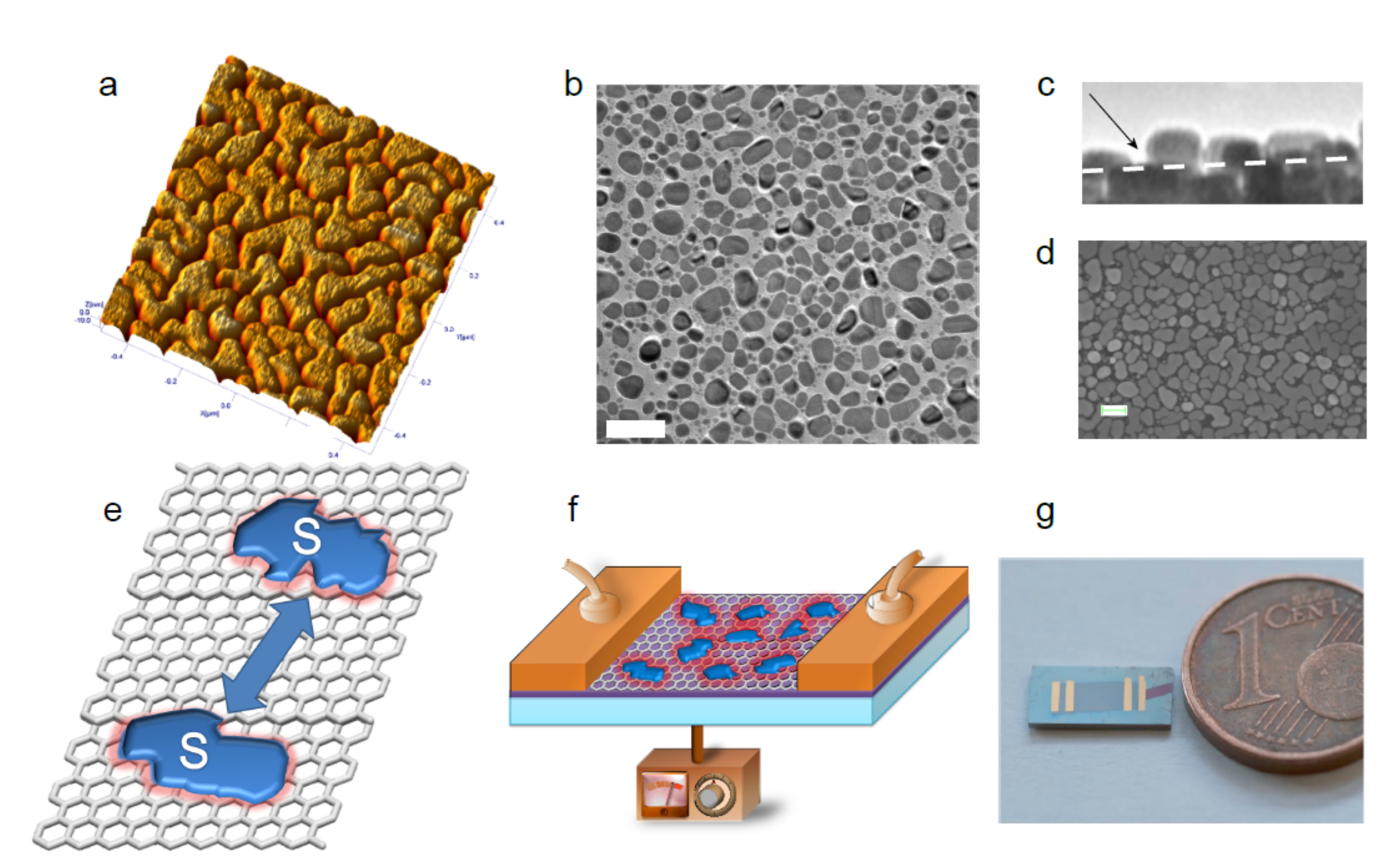}

 \caption{\textbf{Self-assembled graphene/tin nanohybrids.} \textbf{a} : Atomic Force Micrograph of an 1$\mu m^{2}$ area of the device, obtained by dewetting of an evaporated tin film of nominal thickness 10nm. \textbf{b} :  Transmission Electron Micrograph (TEM) of a the decorated sample showing the self-organized network of tin nanoparticles (scale bar 200~nm) separated by clean graphene.  \textbf{c} :  TEM of the sample transferred on a membrane and observed at a grazing angle (the dashed line corresponds to the graphene surface). The negative wetting angle of tin nanoparticles on graphene (black arrow) can be clearly seen. \textbf{d} Scanning Electron Micrograph (SEM) of the Sn nanoparticles network on graphene (scale bar is 300~nm). \textbf{e,f} Sketches of the device. Changing the gate voltage modulates the extension of phase coherent domains in graphene. \textbf{g} Photograph of the studied device. The dark region between the four electrodes is the decorated graphene sheet. The blue cast is due to the presence of tin nanoparticles on the whole surface. The enhanced contrast of the graphene sheet with respect to the silica sides comes from the difference of grain sizes and gaps  between nanoparticles on graphene and on SiO$_2$.  }\label{fig:fig1f}

\end{center}
\end{figure}

The samples, presented in Fig.\ref{fig:fig1f}, were fabricated by connecting CVD-grown graphene flakes transferred on oxidized silicon~\cite{Li2009} (see Methods). A nominal thickness of 10 nm of tin was evaporated on the whole sample by thermal evaporation. Tin's dewetting (see Fig.\ref{fig:fig1f}c) produces a self-assembled, non-percolating array of pancake-like nanoparticles~\cite{Kessler2010} (see Fig.\ref{fig:fig1f}a, b and d). The typical lateral size of a tin island is 80~nm, with a 13~nm gap in-between islands.

Samples have shown two kinds of behavior depending on their room temperature sheet resistance. Devices having the lowest sheet resistance (typically $\leq$~10~k$\Omega$/$\square$) showed superconductivity for all gate voltages with a gate tunable transition temperature, similarly to what was reported for exfoliated samples~\cite{Kessler2010}.  Here, we will focus on the other type of devices, which exhibited a high sheet resistance ($\geq$~15~k$\Omega$/$\square$) at room temperature.

\begin{figure}[htbp]
\begin{center}
	\includegraphics[width=16cm]{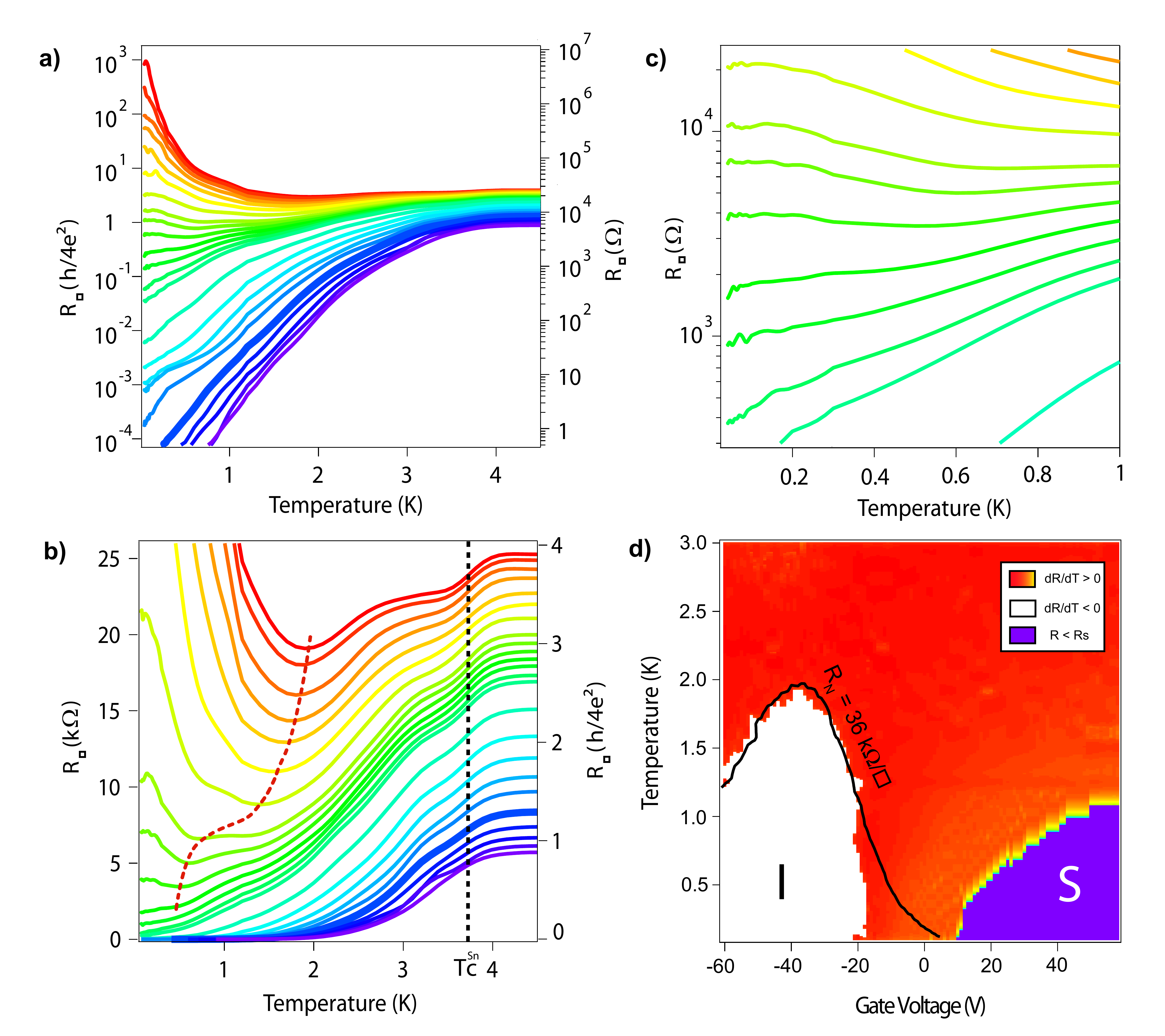}
			 \caption{\textbf{Sheet resistance as a function of temperature for different gate voltages.} \textbf{a} : Sheet resistance as a function of sample temperature for different gate voltages, plotted on a log scale. From top to bottom, voltage offsets from the charge neutrality point ($V_D$~=~-36~V, see Fig.\ref{fig:fig3f}) are $\Delta V~=~Vg~-~V_D$~=~0, 3, 6, 8, 10, 12, 14, 16, 17, 18, 19, 20, 21, 22, 26, 31, 36, 41, 46, 56, 66, 76, 86, 96~V. \textbf{b} : Same data (lower part) plotted on a linear scale to emphasize the behaviour between 1~K and 4~K. The black dashed line indicates the critical temperature of tin. The red dashed line is a guide to the eye showing the minimum resistivity. \textbf{c} : Zoom on the critical region. \textbf{d} : Phase diagram of the superconducting to insulator transition. $\log\left(\frac{\partial R}{\partial T}\right)$ vs gate voltage and temperature. I stands for "Insulating", and S for "Superconducting". We call superconducting the region where $R$ is below the noise floor ($R_S$~=~0.5$\Omega$).  The black line is the iso-value of the normal state resistivity obtained at $B=1$~T (see Supplementary Figure S2) which appeared the closest to the border between the two states.}\label{fig:fig2f}
\end{center}
\end{figure}

\begin{figure}[htbp]
	\includegraphics[width=8cm]{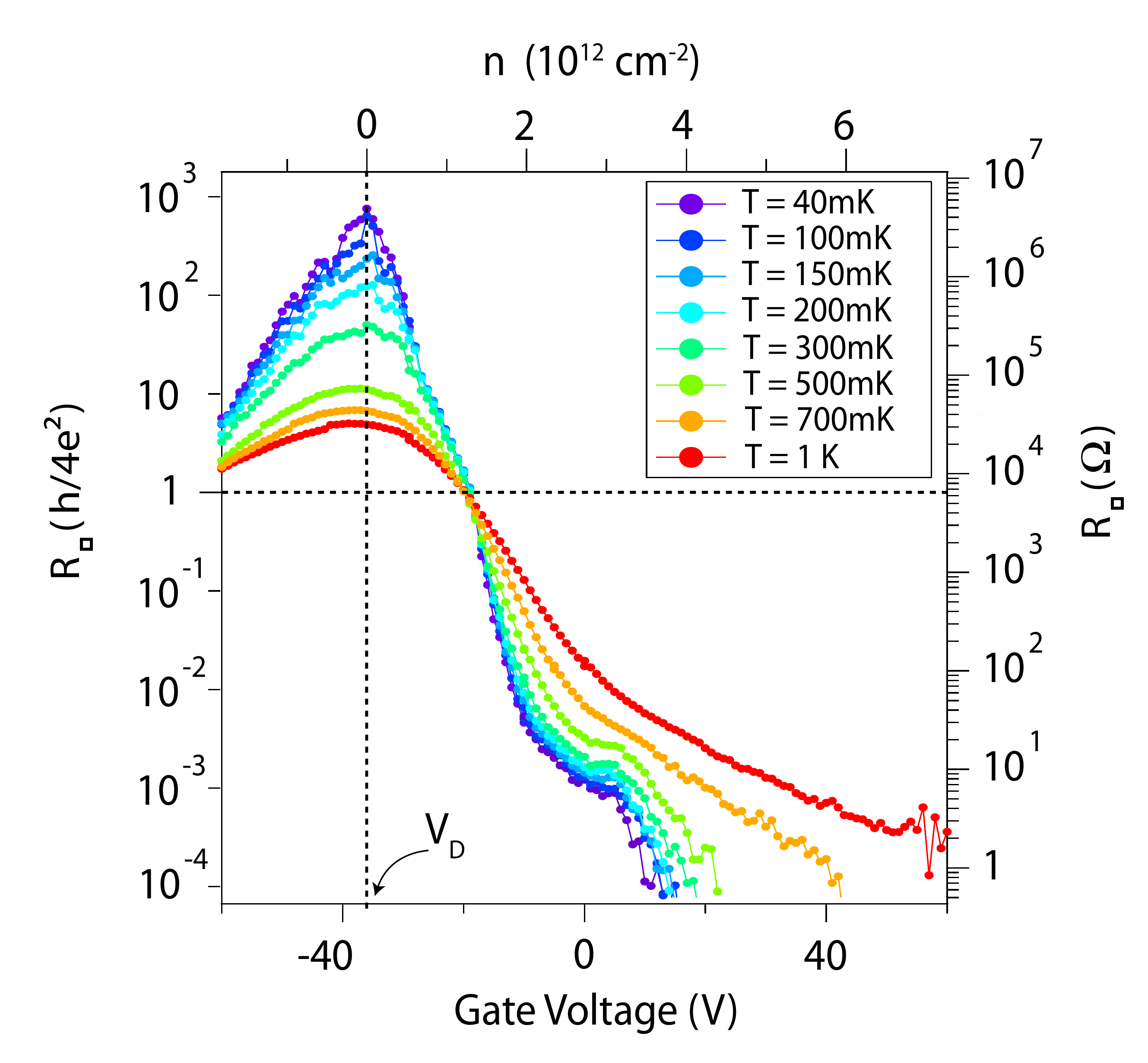}
   \caption{\textbf{Sheet resistance as a function of gate voltage for different temperatures.} Gate voltage dependance of sheet resistance for the lowest temperatures. The vertical dotted line indicated the charge neutrality point. The horizontal one indicated the quantum of resistance for Cooper pairs $R_{\square}$~=~$\frac{h}{4e^2}$ (Top axis : carrier density calculated using the gate capacitance per unit area for 285~nm of SiO$_2$ $C_{bg}$~=~121~$\mu$F/m$^{2}$).}\label{fig:fig3f}
\end{figure}

\bigskip
	Upon cooling from room temperature down to 6K, the resistivity increase ranges from 20\% up to 100\% at the charge neutrality point. This behavior is consistent with the enhanced weak localization and electron-electron interactions expected in 2D metals. Just above the critical temperature of bulk tin ($T_c^{Sn}$~=~3.7~K, black dotted line in Fig.\ref{fig:fig2f}a), a 10\% resistance drop is observed (see  Fig.\ref{fig:fig2f}b), arising from superconducting fluctuations in graphene near the tin islands. Then a broad transition takes place, either towards a superconducting state at high electrostatic doping (for a voltage offset from the charge neutrality point $\Delta V$~>~45~V), or towards an insulating state for voltages closer to the charge neutrality point ($\Delta V$~<~10~V). In between these two gate voltages, the resistance levels off at low temperatures (Fig.\ref{fig:fig2f}c), suggesting an intermediate metallic behavior.
 On the superconducting side, the system follows an "inverse Arrhenius law" $R\propto R_0\exp\left(\frac{T}{T_0}\right)$, as already reported in quench condensed granular films~\cite{Frydman2002}. Significant fluctuations and leveling in the region just before superconductivity sets in (see Fig.\ref{fig:fig3f}, bump near +10~V) are indicative of a percolative behavior. The amplitude of the critical current also supports this picture, as it was repetitively measured to be on the order of 1~$\mu$A, which corresponds to a critical current density of 5.10$^{-4}$~A/m, a value 2,000 times smaller than the value found in samples made from exfoliated graphene~\cite{Kessler2010} or clean CVD graphene.
 
	Above 2K, $R(T)$ curves at all gate voltages behave qualitatively the same. Reducing the temperature further, the curves near the charge neutrality point then reach a minimum at a gate-dependent temperature (red dotted line in Fig.\ref{fig:fig2f}b), below which they start to increase sharply. This re-entrant behavior is reminiscent of what was observed in granular superconductors~\cite{Jaeger1989} or in Josephson junctions arrays~\cite{vanderZant1996}. Our system is indeed similar to a granular superconductor in which the role of the intergranular media is taken up by graphene. In such systems, the SIT is driven by the competition between the charging energy $E_C$ of a superconducting island and the Josephson energy $E_J=\frac{1}{2}\frac{h}{4e^2}\frac{\Delta}{R_N}\tanh\left(\frac{\Delta}{2k_{B}T}\right)$, where $R_N$ is the normal state resistance of the junction. Dissipative degrees of freedom, such as quasiparticles~\cite{Chakravarty1987} or capacitive coupling to a 2DEG~\cite{Rimberg1997} like graphene~\cite{Lutchyn2008}, lead to renormalization of the charging energy. Here, the leveling of resistance in the intermediate 'metallic' regime is indicative of such dissipative processes~\cite{Mason1999}. Dissipation strength scales as $R_N^{-1}$, and $R_N$ in turn depends on gate voltage and temperature, as can be seen by applying a magnetic field above the critical field (see Supplementary Figure S2). 
	$R_N$ thus tunes both energies ($E_J$ and $E_C$) simultaneously, unlike previously considered situations~\cite{Rimberg1997, Lutchyn2008}. Despite this complex dependence, the phase diagram of the SIT (Fig.~\ref{fig:fig2f}d) shows that the boundary of the insulating region (red dotted line in Fig.\ref{fig:fig2f}b) can be related to a constant value of $R_N$.
	Interestingly, when measuring resistance at constant temperature as a function of gate(Fig.\ref{fig:fig3}), one sees a crossing point at $V_G \cong -20$~V where the sheet resistance is of the order of $\frac{h}{(2e)^2}$, the pair quantum of resistance. The universal value of the critical resistance at the transition was predicted by the so-called "dirty bosons model"~\cite{Fisher1990}. Around this transition, the resistance varies by more than 7 orders of magnitude over a gate range of 40~V (corresponding to a carrier density change of 3$\times$10$^{12}$~cm$^{-2}$). This electrostatically driven transition shows a strongly insulating state, with exponential divergence of the resistance.

\bigskip

\begin{figure}[htbp]
	\includegraphics[width=8cm]{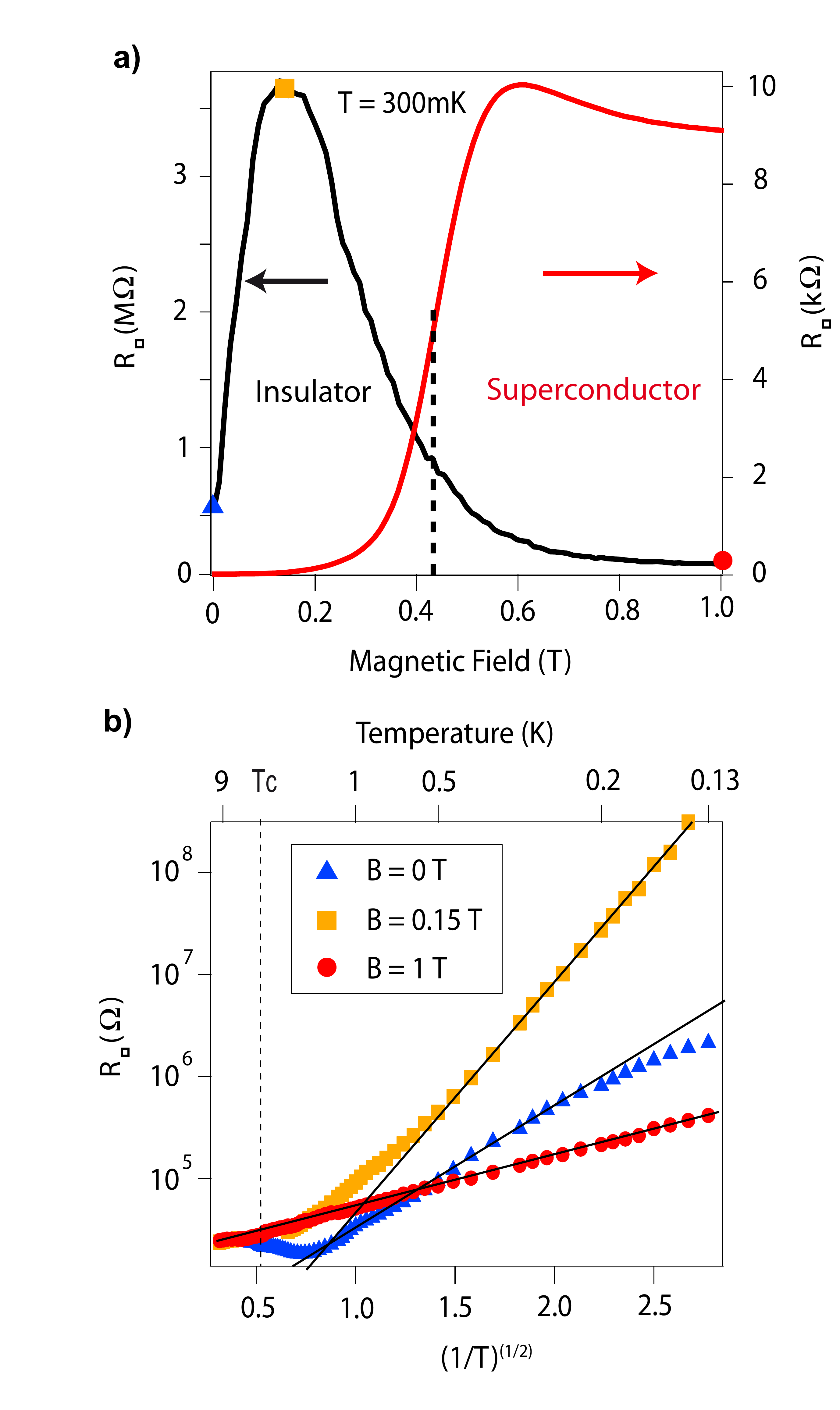}   
\caption{\textbf{Localization of Cooper pairs under a magnetic field.} \textbf{a} : Sheet resistance as a function of magnetic field, measured at $T=300$ mK. Black curve shows the magnetoresistance at the charge neutrality point. The red curve has been measured deep in the superconducting region. \textbf{b} : Temperature dependance of the sheet resistance at three different magnetic fields (indicated in the upper panel). The black lines are fits to the Efros-Schklovsky law, giving the following activation temperatures : $T_1$~=~7.8~K, $T_1'$~=~32.6~K, and $T_1''$~=~2.5~K for $B=0$~T, 0.15~T and 1~T, respectively.}\label{fig:fig4f}
\end{figure}

The magnetoresistance curves in the insulating and superconducting regions are presented in Fig.\ref{fig:fig4f}a and show a peak, both in the superconducting and in the insulating regions. Such a non-monotonic behavior has been widely reported in superconducting thin films. By gating the sample, we observed a continuous crossover between different magnetoresistance regimes~\cite{Steiner2005}. In the superconducting state (red curve in Fig.\ref{fig:fig4f}), the small resistance overshoot at intermediate magnetic field can be understood in terms of Galitski-Larkin correction to the conductivity~\cite{Galitski2001}. The inflection point corresponds to the critical field expected in tin nanoparticles~\cite{Wang2010}. However, the behaviour in the insulating region (black curve in Fig.\ref{fig:fig4f}a) cannot be explained with perturbation theory. Here, the resistance at intermediate field ($B=0.15$~T) is about 40 times higher than the resistance in the normal state ($B=1$~T). Such huge effects have been reported in amorphous thin films~\cite{Baturina2007, NGuyen2009}, and have been explained~\cite{Beloborodov2006} to stem from the underlying nature of superconductivity in amorphous thin films, which is inhomogeneous~\cite{Sacepe2008,Sacepe2011} near the transition. In a granular system, since grains of different sizes have different critical fields, there exists an intermediate field where half of the grains are superconducting (S) and the other half are normal (N). N-S junctions can prevent percolation as they provide barriers for both quasiparticles and Cooper pairs. As we move away from the insulating region (by increasing $\Delta V$), the resistivity maximum is shifted towards higher magnetic fields. This indicates that islands with the smaller critical fields ($B_C$~$\approx$~0.15~T) play a crucial role in the percolation process, whereas deep in the superconducting region, coupling is established directly between other islands. 

 Finally, the temperature dependance of the system when biased at $\Delta V=0$ and for $B=0.15$~T (i.e. in the region where Cooper pairs are localized) does not quite follow the activation law predicted by Beloborodov \textit{et al.}~\cite{Beloborodov2006}. Instead it shows an Efros-Schklovsky like behaviour $T\propto\exp\left(\frac{T_1}{T}\right)^{\frac{1}{2}}$ with an activation energy $T_1$~=~32.6~K (see Fig.\ref{fig:fig4f}b), suggesting that Coulomb interactions may play an important role in the transport.

\begin{figure}[htbp]
	\includegraphics[width=8cm]{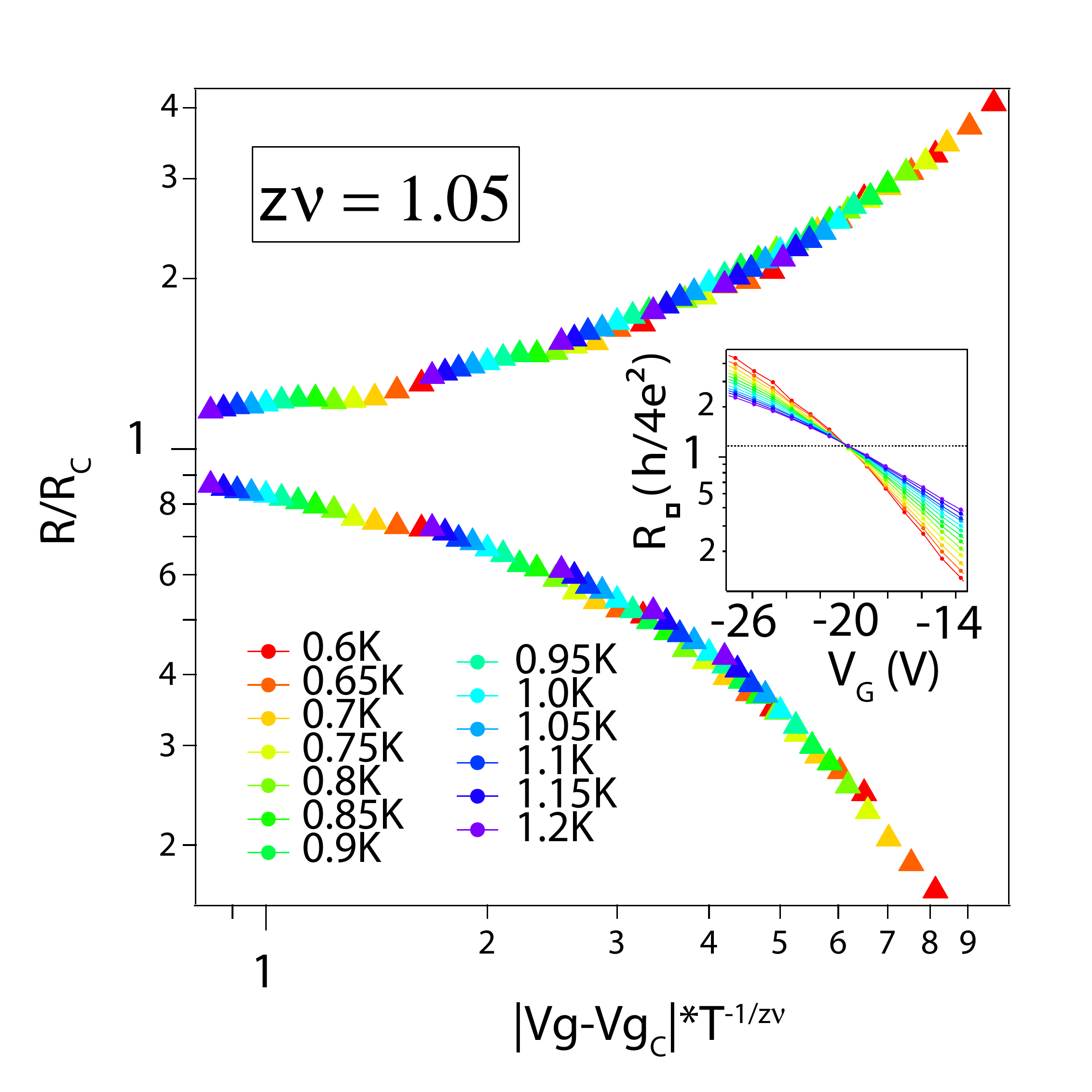}
	     \caption{\textbf{Universal scaling of the transition} Finite-size scaling analysis of the quantum phase transition, using data of Fig.\ref{fig:fig3f}. The best collapse was found using $\nu z$~=~1.05~$\pm$~0.05. Inset : subset of the data used for the scaling : 0.6~K~<~$T$~<~1.3~K and a gate voltage interval of $\pm$~6~V around the critical point $V_{Gc}$~=~-20~V.}\label{fig:fig5f}

\end{figure}

\bigskip
Going back to the gate-induced SIT, Fig.\ref{fig:fig5f} shows how the data can be interpreted as a quantum phase transition~\cite{Sondhi1997}, using finite-size scaling. The voltage range around the critical gate voltage $V_{Gc}$~=~-20~V was chosen the largest possible ($\pm6 V$) while still retaining a universal exponent on both (insulating and superconducting) sides. Below 600 mK, the field effect at the transition shows a significant shift from the universal behavior.  The presence of puddles of normal electrons in the graphene sheet, which are sources of gapless excitations~\cite{Lutchyn2008}, provides a dissipative environment.  The system becomes more coupled at low temperatures to dissipative degrees of freedom, leading to a breakdown of the universal behavior as already observed in experiments involving MoGe films~\cite{Mason1999}. However, above 600 mK, the critical resistance lies very close to the value predicted by the "dirty bosons" model : $\frac{R_C}{R_Q} \cong 1.2$, $R_Q = \frac{h}{4e^{2}}$, which is an indication that we are in the regime of low dissipation, where the dynamical critical exponent $z$ is still equal to 1~\cite{Wagenblast1997}. The exponent $z\nu$ has been evaluated using the two methods described in ~\cite{Markovic1999}. The first method is to multiply each curve by the factor $t$ yielding the best collapse to the first curve, then fitting $t$ to $T^{-z\nu}$. The second method is to take the slope of $\log\left[\left(\frac{\partial R}{\partial Vg}\right)_{V_{g_c}}\right]$ vs $\log\left(T^{-1}\right)$. The first method (shown in Fig.\ref{fig:fig5f}) leads to $z\nu$~=~1.05~$\pm$~0.10 while the second method gives a value of $z\nu$ = 1.18 $\pm$ 0.02 (see Supplementary Figure S4). The second method is probably more accurate, as it is based on data which are taken within the critical region. This value is close to other reported values of $z\nu$ in thickness-tuned transition in Bi~\cite{Markovic1999}. Note that this exponent differs from the expected exponent for classical percolation in 2D ($z\nu$=4/3). It is instead in good agreement with recent theoretical developments on the superfluid transition in disordered two-dimensional bosonic systems~\cite{Refael2011}, which can be understood as the percolation of phase coherent domains into a macroscopic superfluid.

\bigskip
Unlike previously reported gate-induced SITs which either showed a partial SIT transition towards a weakly localized metal~\cite{Parendo2005, Caviglia2008, Bollinger2011} or involved ionic gating, which freezes at low temperatures~\cite{Leng2011}, tin-decorated CVD graphene can be gated continuously at low temperatures with a very strong transconductance (Fig.\ref{fig:fig3f}). This could have application, for example in transition-edge particle detectors. The recently demonstrated metal-insulator transition in ultra-clean graphene samples~\cite{Ponomarenko2011} also opens exciting new perspectives to probe the SIT in the opposite limit of very low disorder. The present experiment paves the way to the realization of more complex graphene-based hybrid materials where graphene acts as a tunable medium or adjustable environment that controls the establishment of long-range electronic orders, such as superconductivity or ferromagnetism. This experiment sheds light on the emergence of superconductivity in inhomogeneous superconductors, and demonstrates the potential of graphene as a versatile substrate for the realization of hybrid superconducting materials.

\bigskip

\section{Methods}

\subsection{Sample Preparation}

Graphene is grown using a chemical vapor deposition (CVD) technique on copper foils (typically 25~$\mu$m thick, from Alfa-Aesar) following the methods described in ~\cite{Li2009}. During the growth, a flow of methane (CH$_4$) provides the carbon feedstock, while forming gas (H$_2$/Ar \NoAutoSpaceBeforeFDP{1:9}) limits the reaction and only a single layer of graphene is obtained. 
After growth, the graphene is protected with a support layer of 1~$\mu$m-thick polymethyl-methacrylate (PMMA), and copper is etched away using a solution of 0.2~g/ml sodium persulfate (Na$_2$S$_2$O$_8$). The graphene remains attached to PMMA and floats in the solution. It is then carefully transferred onto a wafer of degenerately doped oxidized silicon and PMMA is removed using acetone wash followed by thermal annealing at 380°C/1h under Argon atmosphere. 
	Tin is deposited on the whole sample using room temperature Joule evaporation. Pd/Au electrodes are subsequently deposited using a millimeter-scaled metal foil stencil mask in a four-probe geometry aligned on top of the graphene sheet. Supplementary Figure S1 shows a typical sample after fabrication. The fabricated samples were about 5~mm in length and 3~mm in width, sizes that could not previously be obtained using exfoliated graphene. Such a macroscopic sample allowed us to get mesoscopic effects such as universal conductance fluctuations to be averaged out, which is crucial when studying how the phase transition scales. 

Several samples were measured, and the number of graphene layers was varied (from 1 to 3), as well as the thickness of tin (8-20~nm). However, we did not see a direct correlation between these parameters and the behavior of the device. We could only relate it to the normal state sheet resistance of graphene. Only graphene showing a high sheet resistance at room temperature (>15~k$\Omega$) would behave as an insulator below the tin's superconducting transition temperature (3.7~K). The other samples behaved like the ones studied by Kessler et al.~\cite{Kessler2010}, showing much higher critical current density and a gate-tunable Berezinsky-Kozterlitz-Thouless transition towards a superconducting state at all gate voltages. 

\subsection{Measurement setup}
The sample was thermally anchored to the mixing chamber of a $^3$He/$^4$He dilution cryostat and connected to highly filtered measurement lines. The setup allowed the temperature to be continuously varied between 10 K and 0.03K. The differential resistance was recorded using a lock-in amplifier operated in a four-probe configuration at frequencies between 9~Hz and 37~Hz, with an excitation current of 1~nA. In the high impedance state, a two-probe, voltage-biased configuration was used using a Keithley 6430 electrometer or a Femto current-to-voltage converter to record the current.

\section{Acknowledgements}
This work is partially supported by ANR-BLANC SuperGraph, ERC Advanced Grant MolNanoSpin No. 226558 and Cible program from Région Rhone-Alpes. Samples were fabricated at the NANOFAB facility of the Néel Institute, which support team is gratefully acknowledged. We thank  H. Arjmandi-Tash, N. Bendiab, H. Bouchiat, C. Chapelier, J. Coraux, M.V. Feigel'man , Ç.Ö. Girit, B.M. Kessler, L. Marty, A. Reserbat-Plantey, B. Sacépé, V. Sessi, W. Wernsdorfer and A. Zettl for help and stimulating discussions.

\section{Author contributions}
V.B. and A.A conceived the experiments, Z.H. grew the graphene, A.A. and Z.H. fabricated the samples and carried out the measurements, A.A. and V.B. analyzed the data and wrote the paper.

\section{Competing financial interests}
The authors declare no competing financial interests.
\bigskip

\section{Figures}
\newpage

\begin{thebibliography}{breitestes Label}
\bibitem{Geim2007}
Geim, A.K. \& Novoselov, K.S. The rise of graphene. \textit{Nature Mat.} \textbf{6,} 183-191 (2007).
  
 \bibitem{Elias2009}
Elias, D.C. \textit{et al.} Control of graphene's properties by reversible hydrogenation: evidence for graphane. \textit{Science} \textbf{323,} 610-613 (2009).
 
\bibitem{Kozlov2011}
Kozlov, S.M., Vines; F. \& Görling, A. Bandgap engineering of graphene by physisorbed adsorbates. \textit{Adv. Mat.} \textbf{23,} 2638-2643 (2011).

\bibitem{Ponomarenko2011}
Ponomarenko, A.L. \textit{et al.} Tunable metal-insulator transition in double-layer graphene heterostructure. \textit{Nature Phys.},\textbf{7,}, 958-961 (2011).

\bibitem{Candini2011} 
Candini, A., \textit{et al.} Graphene Spintronic Devices with Molecular Nanomagnets, \textit{Nano Lett.} \textbf{11,} 2634-2639 (2011).

\bibitem{Uchoa2007}
Uchoa, B.,  \& Castro Neto, A.-H., Superconducting States of Pure and Doped Graphene, \textit{Phys. Rev. Lett.} \textbf{98,} 146801 (2007).

\bibitem{Heersche2007}
Heersche, H.B., Jarillo-Herrero, P., Oostinga, J.B., Vandersypen, L. \& Morpurgo, A.F. Bipolar supercurrent in graphene. \textit{Nature} \textbf{446,} 56-59 (2007).

\bibitem{Feigel'man2008}
Feigel'man, M.V., Skvortsov, M.A. \& Tikhonov, K.S. Proximity-induced superconductivity in graphene. \textit{JEPT Lett.} \textbf{88,} 747-751 (2008).

\bibitem{Kessler2010}
Kessler, B.M., Girit, C,Ö, Zettl, A. \& Bouchiat, V. Tunable superconducting phase transition in metal-decorated graphene sheets. \textit{Phys. Rev. Lett.} \textbf{104,} 047001 (2010).

\bibitem{Goldman1998}
Goldman, A.M. \& Markovic, N. Superconductor-insulator transitions in the two-dimensional limit. \textit{Phys. Today} \textbf{51,} No. 11, 39-43 (1998). 

\bibitem{Sondhi1997}
Sondhi, S.L., Girvin, S.M., Carini, J.P. \& Shahar, D. Continuous quantum phase transitions. \textit{Rev. Mod. Phys.} \textbf{69,} 315 (1997).

\bibitem{Bollinger2011}
Bollinger, A.T. \textit{et al.} Superconductor-insulator transition in La$_{2x}$Sr$_x$CuO$_4$ at the pair quantum resistance. \textit{Nature} \textbf{472,} 458-460 (2011).

\bibitem{Parendo2005}
Parendo, K.A. \textit{et al.} Electrostatic tuning of the superconductor-insulator transition in two dimensions. \textit{Phys. Rev. Lett.} \textbf{94,} 197004 (2005).

\bibitem{Leng2011}
Leng, X., Garcia-Barriocanal, J., Bose, S., Lee, Y. \& Goldman, A.M. Electrostatic control of the evolution from a superconducting phase to an insulating phase in ultrathin YBa$_2$Cu$_3$O$_7-x$ films. \textit{Phys. Rev. Lett.} \textbf{107,} 027001 (2011).

\bibitem{Li2009}
Li, X.S. \textit{et al.}, Large-area synthesis of high-quality and uniform graphene films on copper foils. \textit{Science} \textbf{324,} 1312-1314 (2009).

\bibitem{Frydman2002}
Frydman, A., Naaman, O. \& Dynes, R.C., Universal transport in two-dimensional granular superconductors. \textit{Phys. Rev. B} \textbf{66,} 052509 (2002). 

\bibitem{Jaeger1989}
Jaeger, H.M., Haviland, D.B., Orr, B.G. \& Goldman, A.M. Onset of superconductivity in ultrathin granular metal films. \textit{Phys. Rev. B} \textbf{40,} 182 (1989).

\bibitem{vanderZant1996}
van der Zant, H.S.J., Elion, W.J., Geerlings, L.J. \& Mooij, J.E., Quantum phase transitions in two dimensions: experiments in Josephson-junction arrays. \textit{et al.}, Phys. Rev. B \textbf{54}, 10081 (1996).

\bibitem{Chakravarty1987}
Chakravarty, S., Kivelson, S., Zimanyi, G.T. \& Halperin, B.I., Effect of quasiparticle tunneling on quantum-phase fluctuations and the onset of superconductivity in granular films. \textit{Phys. Rev. B} \textbf{35,} 7256 (1987).

\bibitem{Rimberg1997}
Rimberg, A.J. \textit{et al.}, Dissipation-driven superconductor-insulator transition in a two-dimensional Josephson-junction array. \textit{Phys. Rev. Lett.} \textbf{78,} 2632 (1997).

\bibitem{Lutchyn2008}
Lutchyn, R.M., Galitski, V., Refael, G. \& Das Sarma, S. Dissipation-driven quantum phase transition in superconductor-graphene systems. \textit{Phys. Rev. Lett.} \textbf{101,} 106402 (2008).

\bibitem{Mason1999}
Mason, N. \& Kapitulnik, A., Dissipation effects on the superconductor-insulator transition in 2D superconductors. \textit{Phys. Rev. Lett.} \textbf{82,} 5341 (1999).

\bibitem{Fisher1990}
Fisher, M.P.A., Grinstein, G. \& Girvin, S.M. Presence of quantum diffusion in two dimensions: Universal resistance at the superconductor-insulator transition. \textit{Phys. Rev. Lett.} \textbf{64,} 587 (1990).


\bibitem{Steiner2005}
Steiner, M.A., Boebinger, G. \& Kapitulnik, A., Possible field-tuned superconductor-insulator transition in high-Tc superconductors: implications for pairing at high magnetic fields. \textit{Phys. Rev. Lett.} \textbf{94,} 107008 (2005).

\bibitem{Galitski2001}
Galitski, V.M. \& Larkin, A.I., Superconducting fluctuations at low temperature. \textit{Phys. Rev. B} \textbf{63,} 174506 (2001).

\bibitem{Wang2010}
Wang, X-L., Feygenson, M., Aronson, M.C. \& Han, W.-Q., Sn/SnOx core-shell nanospheres: synthesis, anode performance in Li-Ion batteries, and superconductivity. \textit{J. Phys. Chem. C} \textbf{114,} 14697-14703 (2010). 

\bibitem{NGuyen2009}
Nguyen, H.Q., \textit{et al.}, Observation of giant positive magnetoresistance in a Cooper pair insulator. \textit{Phys. Rev. Lett.} \textbf{103,} 157001 (2009).

\bibitem{Baturina2007}
Baturina, T.I. , Mironov, A.Yu., Vinokur, V.M., Baklanov, M.R. \& Strunk, C., Localized superconductivity in the quantum-critical region of the disorder-driven superconductor-insulator transition in TiN thin films. \textit{Phys. Rev. Lett.} \textbf{99,} 257003 (2007).


\bibitem{Beloborodov2006}
Beloborodov, I.S., Fominov, Ya.V., Lopatin, A.V. \& Vinokur, V.M., Insulating state of granular superconductors in a strong-coupling regime. \textit{Phys. Rev. B} \textbf{74,} 014502 (2006).

\bibitem{Sacepe2008}
Sacépé, B. \textit{et al.} Disorder-induced inhomogeneities of the superconducting state close to the superconductor-insulator transition. \textit{Phys. Rev. Lett.} \textbf{101,} 157006 (2008).

\bibitem{Sacepe2011}
Sacépé, B. \textit{et al.}, Localization of preformed Cooper pairs in disordered superconductors. \textit{Nature Phys.} \textbf{7,} 239-244 (2011).

\bibitem{Wagenblast1997}
Wagenblast, K.-H., van Otterlo, A., Schön, G. \& Zimanyi, G.T., New universality class at the superconductor-insulator transition. \textit{Phys. Rev. Lett.} \textbf{78,} 1779 (1997).

\bibitem{Markovic1999}
Markovic, N., Christiansen, C., Mack, A.M., Huber, W.H. \& Goldman, A.M., Superconductor-insulator transition in two dimensions. \textit{Phys. Rev. B} \textbf{60,} 4320 (1999).  


\bibitem {Refael2011}
Iyer, S., Pekker, D. \& Refael, G. A Mott glass to superfluid transition for random bosons in two dimensions. \textit{Phys. Rev. B} \textit{in press}  (2012), available at arXiv:1110:3338v2.



\bibitem{Caviglia2008}
Caviglia, A.D. \textit{et al.} Electric field control of the LaAlO3/SrTiO3 interface ground state. \textit{Nature} \textbf{456,} 624-627 (2008).

\end{thebibliography}
\end{document}